\documentclass[aps,twocolumn,showpacs,pra]{revtex4}
\usepackage{amsmath}
\usepackage{amsfonts}
\usepackage{amssymb}
\usepackage{epsfig}
\usepackage{multirow}
\usepackage{graphicx}
\begin{document}
\title{Robust force sensing for a free particle in a dissipative optomechanical system with a parametric amplifier}
\author{Sumei Huang$^{1}$ and G. S. Agarwal$^{2}$}
\affiliation{$^1$ Room 105, Building 4, Erqu, Fuzhou Erhua Xincun, Fuzhou, Fujian 350011, China}
\affiliation{$^2$ Institute for Quantum Science and Engineering and Department of Biological and Agricultural Engineering,
Texas A\&M University, College Station, Texas 77845, USA}
\date{\today}
\begin{abstract}

We theoretically investigate optical detection of a weak classical force acting on a free particle
in a dissipative coupling optomechanical system with a degenerate parametric amplifier (PA). We
show that the PA allows one to achieve the force sensitivity far better than the standard quantum limit (SQL) over a broad range of the detection frequencies. The improvement depends on the parametric gain and the driving power. Moreover, we discuss the effects of the mechanical damping
and the thermal noise on the force sensitivity. We find that the robustness of the force sensitivity much
better than the SQL against the mechanical damping and the thermal noise is achievable in the
presence of the PA with a high parametric gain. For the temperature $T = 1$ K, the improvement in
sensitivity is better by a factor of about 7 when the driving power is set at a value corresponding to the SQL with no PA.
\end{abstract}
\maketitle
\section{Introduction}
The dissipative optomechanical coupling systems have recently attracted considerable attention. The dissipative coupling is characterized by the dependence of the cavity decay rate on the displacement of the mechanical oscillator \cite{Clerk}. Several dissipative optomechanical systems have been demonstrated experimentally, these include a microdisk resonator coupled to a vibrating nanowaveguide \cite{Li}, a Michelson--Sagnac interferometer containing a moving membrane \cite{Hammerer3}, a photonic crystal split-beam nanocavity \cite{Wu1, Wu2}. The dissipative coupling offers several advantages, for example, it has been shown that the dissipative coupling can lead to near ground-state cooling of the mechanical resonator in the unresolved sideband limit \cite{Clerk,Hammerer1,Hammerer2,Hammerer3,Weiss,Wu1} and the squeezing of the mechanical oscillator \cite{Huang1, Li1, Li2}. A variety of other physical effects with dissipative coupling have been discussed: the normal mode splitting \cite{Huang2,Nuunenkamp}, the electromagnetically induced transparency \cite{Nuunenkamp}, the squeezing of the output light \cite{Qu, Kilda}, and many others \cite{Liew,Sete}. An important application of the dissipative coupling is in the sensitive detection of the force and torque \cite{Matsko,Wu2}. Recently, the dissipative optomechanical coupling has been shown to overcome the standard quantum limit (SQL) \cite{Matsko}. It has been found that the sensitivity for measurement of a classical signal force acting on a free particle in a dissipative optomechanical system is comparable with that in a dispersive optomechanical system but has much wider measurement bandwidth \cite{Matsko}.

 In an early paper the present authors had demonstrated the great advantage of using a parametric amplifier (PA) in the optomechanical cavity \cite{cooling}. For example, it was shown that the dispersive optomechanical coupling systems with parametric amplification can aid in cooling the movable mirror \cite{cooling} and in observing the normal-mode splitting in the coupling of the movable mirror and the output field \cite{NMS}. Our original suggestion \cite{cooling} has been followed by many others \cite{Xuereb,Peano,Squeezing,Lu}.

 Guided by these wide range of new features when the PA is in the cavity, we study the dissipative optomechanical system with a degenerate PA. It is found that the PA greatly enlarges the range of detection frequencies in which the force sensitivity is better than the SQL, and largely improve the force sensitivity.
We further study the effects of the mechanical damping and the mechanical thermal noise on the force sensitivity in the absence and presence of the PA. We show that the detrimental effects of the mechanical damping and the thermal noise on the force sensitivity can be significantly suppressed in the presence of the PA with a high parametric gain. For the temperature $T=1$ K, the improvement in the force sensitivity can be by a factor of about 7 \cite{fn}.

It is to be noted that the detection force using the dispersive optomechanical system has a long history with many pioneering contributions.
To realize high sensitive force measurements, many approaches have been proposed to reduce quantum noise and surpass the SQL in the force measurement. It has been shown that the force sensitivity can be improved by using a squeezed vacuum \cite{Caves2,Kimble}, optomechanical velocity measurement \cite{VF, VMFK}, Kerr media \cite{Bondurant}, a dual-mechanical-resonator scheme \cite{Briant,Caniard}, a signal-recycling mirror \cite{Chen1,Chen2}, two-tone drives \cite{Clerk08, Suh}, a second, auxiliary optical cavity coupled to
the optomechanical cavity \cite{Tsang1, Tsang2, Hammerer}, a mirror-in-the-middle optomechanical system with two coupled cavity modes \cite{Xu}. There are also proposals combining optomechanical systems with ultracold atoms \cite{Seok,Vitali}. However, on the experimental side using ultracold atoms the SQL is still not beaten \cite{Science}.

The paper is organized as follows. In Sec. II, we introduce the model, give the quantum Langevin equations and the
steady-state mean values, calculate the spectrum of fluctuations in the generalized quadrature of the output light, give the analytic expression for the force sensitivity, find the optimal quadrature angle for the maximal force sensitivity. Then we analyze how the PA in the dissipative system affects the force sensitivity. In Sec. III, we study the influences of the mechanical damping and the thermal noise on the force sensitivity with and without a PA in the dissipative system. In Sec. IV, we summarize our main conclusions.

\section{Model-Parametric amplifier in an optomechanical cavity with a particle of mass $m$}
We consider a degenerate PA in an optomechanical system \cite{Matsko} where a free particle with mass $m$ is dissipatively coupled to a cavity field $c$ with resonance frequency $\omega_{c}$. The cavity field is driven by a strong coherent light with frequency $\omega_{l}$ and amplitude $\varepsilon_{l}$. The parametric amplifier is coupled to the cavity mode. There is no parametric drive on the free particle. For a free particle, the potential energy is zero, its momentum and kinetic energy are conserved. In the degenerate PA, a pump field with a frequency of 2$\omega_{l}$ interacts with a second-order nonlinear optical crystal, thus the output of the PA is at frequency $\omega_{l}$. The Hamiltonian of the system in the rotating frame at the input laser frequency $\omega_{l}$ reads
\begin{eqnarray}\label{1}
H&=&\hbar(\omega_c-\omega_l)c^{\dag}c+\frac{p^{2}}{2m}+i\hbar\sqrt{2\kappa(q)}[\varepsilon_{l}(c^{\dag}-c)\nonumber\\
& &+c^{\dag}c_{in}-c_{in}^{\dag}c]+i\hbar G(c^{\dag2}-c^{2}),
\end{eqnarray}
where the photon decay rate is dependent on the displacement $q$ of the free particle, denoted by $\kappa(q)$, $p$ is the momentum of the free particle, $q$ and $p$ operators satisfy the commutation relation $[q,p]=i\hbar$, $\varepsilon_{l}$ is dependent on the laser power $\wp$ by $\varepsilon_{l}=\sqrt{\frac{\wp}{\hbar\omega_{l}}}$. The $c_{in}$ is the input vacuum noise with zero mean value. The $G$ is the parametric gain of the PA, which is proportional to the pump driving the PA. If the displacement $q$ of the free particle is very small, $\kappa(q)\approx\kappa_{0}+\kappa_{cf}q=\kappa_{0}(1+\eta q)$, where $\kappa_{0}$ is the photon decay rate for $q=0$, $\kappa_{cf}$ is the dissipative coupling constant between the cavity field and the free particle, and $\eta=\kappa_{cf}/\kappa_0$. The parameter $\eta$ using the realizable system of a Michelson--Sagnac interferometer with a silicon
nitride membrane \cite{Hammerer3} is $4.182\times10^{8}\ \frac{1}{\text{m}}$ \cite{fn1}. Thus $\sqrt{2\kappa(q)}\approx\sqrt{2\kappa_{0}}(1+\frac{\eta}{2}q)$. In Eq. (\ref{1}), the first two term are the free energies of the cavity field and the free particle, respectively, the third term describes the couplings of the cavity field with an external laser and the input vacuum noise $c_{in}$, the last term shows the interaction between the cavity field and the PA. In the following, we consider the case that the cavity mode is resonantly pumped by the input laser
 ($\omega_{c}=\omega_{l}$). If there is an external weak force $f_{ex}$ with zero mean value acting on the free particle, we use Heisenberg's equation of motion to find the equations of motion for the system operators. The equations of motion are given by
\begin{eqnarray}\label{2}
\dot{q}&=&\frac{p}{m},\nonumber\\
\dot{p}&=&-\sqrt{2\kappa_{0}}\frac{\eta}{2}i\hbar[\varepsilon_{l}(c^{\dag}-c)+c^{\dag}c_{in}-c_{in}^{\dag}c]+f_{ex},\nonumber\\
\dot{c}&=&\sqrt{2\kappa_{0}}(1+\frac{\eta}{2}q)(\varepsilon_{l}+c_{in})+2Gc^{\dag}-\kappa_{0}(1+\eta q)c.
\end{eqnarray}
 We assume that the steady-state displacement of the free particle is $q_{s}=0$. In the steady state, $\dot{q}=0$, $\dot{p}=0$, $\dot{c}=0$, we find
the steady-state mean value $p_{s}$ of the momentum of the free particle and the steady-state amplitude $c_{s}$ of the cavity field
\begin{eqnarray}\label{3}
p_{s}=0, \quad c_{s}=\frac{\sqrt{2\kappa_{0}}\varepsilon_{l}}{\kappa_{0}-2G}.
\end{eqnarray}
We
have to require $G<\frac{\kappa_{0}}{2}$ to make sure that the system is stable. Note that the equations of motion are nonlinear, which can be linearized in the strong driving regime $|c_{s}|\gg1$. By writing each operator in Eq. (\ref{2}) as $q=q_{s}+\delta q$, $p=p_{s}+\delta p$, $c=c_{s}+\delta c$, where $\delta q$, $\delta p$, $\delta c$ are the small fluctuation operators with zero mean values, keeping the first order in the fluctuations, we obtain the linearized equations for the fluctuating operators
\begin{eqnarray}\label{4}
\delta\dot{q}&=&\frac{\delta p}{m},\nonumber\\
\delta\dot{p}&=&-\sqrt{2\kappa_{0}}\frac{\eta}{2}i\hbar[\varepsilon_{l}(\delta c^{\dag}-\delta c)+c_{s}(c_{in}-c_{in}^{\dag})]+f_{ex},\nonumber\\
\delta\dot{c}&=&-\kappa_{0}\delta c-(\kappa_{0}+2G)\frac{\eta}{2}c_{s}\delta q+2G\delta c^{\dag}+\sqrt{2\kappa_{0}}c_{in}.\nonumber\\
\end{eqnarray}
\subsection{Fluctuations in the output field}
By taking a
Fourier transform $f(t)=\frac{1}{2\pi}\int^{+\infty}_{-\infty} f(\omega)e^{-i\omega t}d\omega$ of all the operators and noise sources in Eq. (\ref{4}), we obtain the fluctuation of the cavity field in the frequency domain
\begin{eqnarray}\label{5}
& &\delta c(\omega)\nonumber\\
&=&\frac{1}{(\kappa_{0}-i\omega)^{2}-4G^{2}}\Big[-(\kappa_{0}-i\omega+2G)(\kappa_{0}+2G)\frac{\eta}{2}c_{s}\nonumber\\
& &\times\delta q(\omega)+\sqrt{2\kappa_{0}}(\kappa_{0}-i\omega)c_{in}(\omega)+2G\sqrt{2\kappa_{0}}c_{in}^{\dag}(-\omega)\Big],\nonumber\\
\end{eqnarray}
where the position fluctuation of the free particle is given by
\begin{eqnarray}\label{6}
\delta q(\omega)&=&\frac{i\hbar\eta\sqrt{\kappa_{0}}c_{s}}{\sqrt{2}m\omega^{2}}\frac{4G-i\omega}{\kappa_{0}-i\omega+2G}[c_{in}(\omega)-c_{in}^{\dag}(-\omega)]\nonumber\\
& &+\frac{f_{ex}(\omega)}{-m\omega^{2}}.
\end{eqnarray}
The optical output field is related to the input
field via the standard input-output relation \cite{Walls}
\begin{eqnarray}\label{7}
c_{out}&=&\sqrt{2\kappa(q)}c-c_{in},
\end{eqnarray}
the fluctuation of the output field can be written as
\begin{eqnarray}\label{8}
\delta c_{out}(\omega)&=&\sqrt{2\kappa_{0}}\delta c(\omega)+\sqrt{2\kappa_{0}}\frac{\eta}{2}c_{s}\delta q(\omega)-c_{in}(\omega).
\end{eqnarray}
Substituting $\delta c(\omega)$ into $\delta c_{out}(\omega)$, we have
\begin{eqnarray}\label{9}
\delta c_{out}(\omega)&=&\sqrt{2\kappa_{0}}\frac{\eta}{2}c_{s}\frac{-i\omega-4G}{\kappa_{0}-i\omega-2G}\delta q(\omega)\nonumber\\
& &+\Big[\frac{2\kappa_{0}(\kappa_{0}-i\omega)}{(\kappa_{0}-i\omega)^{2}-4G^{2}}-1\Big]c_{in}(\omega)\nonumber\\
& &+\frac{4G\kappa_{0}}{(\kappa_{0}-i\omega)^{2}-4G^{2}}c_{in}^{\dag}(-\omega).
\end{eqnarray}
We note that in Ref. \cite{Matsko}, the authors restrict the detection range of frequencies to those much smaller than $2\kappa_0$. This is because they were discussing a direct probe of the velocity of the free particle. We are discussing the force sensitivity which is related to the time derivative of the momentum. We do not need such a condition.
For convenience, we introduce the amplitude and phase quadratures of the input vacuum noise as $x_{in}=\frac{1}{\sqrt{2}}(c_{in}+c_{in}^{\dag})$ and $y_{in}=\frac{1}{i\sqrt{2}}(c_{in}-c_{in}^{\dag})$, and introduce the amplitude and phase quadratures of the output field as $\delta x_{out}=\frac{1}{\sqrt{2}}(\delta c_{out}+\delta c_{out}^{\dag})$ and $\delta y_{out}=\frac{1}{i\sqrt{2}}(\delta c_{out}-\delta c_{out}^{\dag})$.
The amplitude quadrature of the output field is found to be
\begin{eqnarray}\label{10}
\delta x_{out}(\omega)&=&\frac{1}{\sqrt{2}}[\delta c_{out}(\omega)+\delta c_{out}^{\dag}(-\omega)]\nonumber\\
&=&\frac{\kappa_{0}+2G+i\omega}{\kappa_{0}-2G-i\omega}\Big[x_{in}(\omega)+\mathcal{K}(\omega)y_{in}(\omega)\nonumber\\
& &+\sqrt{2\mathcal{K}(\omega)}u(\omega)\frac{f_{ex}(\omega)}{F_{SQL}(\omega)}\Big],
\end{eqnarray}
where
\begin{eqnarray}\label{11}
\mathcal{K}(\omega)&=&\mathcal{J}\frac{\kappa_{0}^{2}}{(\kappa_{0}+2G)^{2}+\omega^{2}}\frac{\omega^{2}+16G^{2}}{\omega^{2}},\nonumber\\
\mathcal{J}&=&\frac{\hbar\eta^{2}c_{s}^{2}}{m\kappa_{0}},\nonumber\\
u(\omega)&=&\frac{\sqrt{16G^{2}+\omega^{2}}}{\sqrt{(\kappa_{0}+2G)^{2}+\omega^{2}}}\frac{\kappa_{0}+2G-i\omega}{\omega+4Gi}i,\nonumber\\
F_{SQL}(\omega)&=&\sqrt{2m\hbar\omega^{2}},
\end{eqnarray}
where $\mathcal{K}(\omega)$ and $\mathcal{J}$ are dimensionless, $\mathcal{J}$ is proportional to the power $\wp$ of the input laser, $F_{SQL}(\omega)$ is the SQL for a free particle's sensitivity to a weak classical force \cite{Matsko}, its dimension is NHz$^{-1/2}$. In Eq. (\ref{10}), the first term is the photon shot noise from the
amplitude quadrature of the input vacuum noise, the second term is the radiation backaction noise from the phase quadrature of the input vacuum noise which is proportional to the input laser power, the third term is from the external weak force.
Moreover, the phase quadrature of the output field is found to be
\begin{eqnarray}\label{12}
\delta y_{out}(\omega)&=&\frac{1}{i\sqrt{2}}[\delta c_{out}(\omega)-\delta c_{out}^{\dag}(-\omega)]\nonumber\\
&=&\frac{\kappa_{0}-2G+i\omega}{\kappa_{0}+2G-i\omega}y_{in}(\omega).
\end{eqnarray}
It is found that the phase quadrature of the output field only depends on the phase quadrature of the input vacuum noise. In the absence of the PA ($G=0$), Eqs. (\ref{10}) and (\ref{12}) are the same as the results in \cite{Matsko}. We define an arbitrary quadrature of the output field as $\delta z_{out}(\omega)=\delta x_{out}(\omega)\cos\phi+\delta y_{out}(\omega)\sin\phi$ with $\phi$ being the homodyne phase angle determined by
the local oscillator. Through calculations, we find
\begin{eqnarray}\label{13}
\delta z_{out}(\omega)&=&\frac{\kappa_{0}+2G+i\omega}{\kappa_{0}-2G-i\omega}\Big\{x_{in}(\omega)\cos\phi\nonumber\\
& &+\Big[\mathcal{K}(\omega)\cos\phi+\frac{1}{A(\omega)}\sin\phi\Big]y_{in}(\omega)\nonumber\\
& &+\sqrt{2\mathcal{K}(\omega)}u(\omega)\frac{f_{ex}(\omega)}{F_{SQL}(\omega)}\cos\phi\Big\},
\end{eqnarray}
where $A(\omega)=\frac{(\kappa_{0}+2G)^{2}+\omega^{2}}{(\kappa_{0}-2G)^{2}+\omega^{2}}$. Here the quadrature $\delta z_{out}(\omega)$ depends on the amplitude quadrature of the input vacuum noise, the phase quadrature of the input vacuum noise, and the external force. Hence $\delta z_{out}(\omega)$ contains information on
the weak force. So the weak force can be detected by performing homodyne detection of the quadrature $\delta z_{out}(\omega)$ of the output field
from the cavity \cite{Walls}.
We define the spectrum of fluctuations in the quadrature $\delta z_{out}(\omega)$ of the output light as
\begin{eqnarray}\label{14}
& &\frac{1}{2}[\langle \delta z_{out}(\omega)\delta z_{out}(\Omega)\rangle+\langle \delta z_{out}(\Omega)\delta z_{out}(\omega)\rangle]\nonumber\\
&=&2\pi S_{zout}(\omega)\delta (\omega+\Omega).
\end{eqnarray}
With the help of the correlation functions of the input vacuum noise
\begin{eqnarray}\label{15}
\langle x_{in}(\omega)x_{in}(\Omega)\rangle&=&\langle y_{in}(\omega)y_{in}(\Omega)\rangle=\frac{1}{2}2\pi\delta(\omega+\Omega),\nonumber\\
\langle x_{in}(\omega)y_{in}(\Omega)\rangle&=-&\langle y_{in}(\omega)x_{in}(\Omega)\rangle=\frac{i}{2}2\pi\delta(\omega+\Omega),\nonumber\\
\end{eqnarray}
we obtain the following expression for the spectrum of the output light
\begin{eqnarray}\label{16}
S_{zout}(\omega)&=&A(\omega)\Big\{\frac{1}{2}\cos^{2}\phi+\frac{1}{2}\Big[\mathcal{K}(\omega)\cos\phi+\frac{1}{A(\omega)}\sin\phi\Big]^{2}\nonumber\\
& &+2\mathcal{K}(\omega)\cos^{2}\phi\frac{S_{ex}(\omega)}{F_{SQL}^{2}(\omega)}\Big\},
\end{eqnarray}
where $2\pi S_{ex}(\omega)\delta(\omega+\Omega)=\langle f_{ex}(\omega)f_{ex}(\Omega)\rangle$, and $F_{SQL}^{2}(\omega)$ is the spectral density of the SQL for force at frequency $\omega$.

\subsection{Force sensitivity beyond the SQL with the PA}
 In this section, we demonstrate how the use of the PA in the cavity can improve the force sensitivity far beyond the SQL for operating parameters for which the sensitivity is at the SQL in the absence of the PA.
We denote the spectrum of the output field without the external force $f_{ex}$ acting on the free particle as $S_{flcut}(\omega)$, then the force sensitivity of the system is determined by the quantity
\begin{eqnarray}\label{17}
R(\omega)&=&\frac{S_{fluct}(\omega)}{\frac{\partial S_{zout}(\omega)}{\partial S_{ex}(\omega)}}\nonumber\\
&=&F_{SQL}^{2}(\omega)\frac{1}{4\mathcal{K}(\omega)}\left\{1+\Big[\mathcal{K}(\omega)+\frac{1}{A(\omega)}\tan\phi\Big]^{2}\right\}.\nonumber\\
\end{eqnarray}
The force sensitivity $R(\omega)$ has two contributions: the first term is the contribution of the photon shot noise from the amplitude quadrature of the input vacuum noise, which is inversely proportional to the laser power $\wp$, the second term is the contribution of the radiation backaction noise from the phase quadrature of the input vacuum noise. We give a comparison of the force sensitivity of the mechanical oscillator with the free particle in Appendix, it is found that two agree well if the detection frequency is much bigger than the mechanical frequency $\omega_{m}$, which is much smaller than $\kappa_{0}$.
 The dimension of $R(\omega)$ is N$^{2}$Hz$^{-1}$. If the dimensionless quantity $\frac{R(\omega)}{F_{SQL}^{2}(\omega)}$ is less than $\frac{1}{2}$, which is the spectral density of the vacuum state (SQL), the SQL is beaten.
 We optimize the homodyne phase $\phi$ so as to minimize
$R(\omega)$. When the second term in $R(\omega)$ is zero, we obtain the optimal phase
\begin{eqnarray}\label{18}
\tan\phi^{opt}=-A(\omega)\mathcal{K}(\omega),
\end{eqnarray}
the minimum value of $R(\omega)$ is found to be
\begin{eqnarray}\label{19}
R(\omega)_{min}=\frac{F_{SQL}^{2}(\omega)}{4\mathcal{K}(\omega)},
\end{eqnarray}
which corresponds to the maximal force sensitivity. For convenience, we define
\begin{eqnarray}\label{20}
\mu(\omega)=\frac{R(\omega)_{min}}{F_{SQL}^{2}(\omega)}=\frac{1}{4\mathcal{K}(\omega)},
\end{eqnarray}
which is dimensionless.
Note that $\mu(\omega)$ is only dependent on the photon shot noise.
 Hence, by choosing optimal homodyne phase $\phi^{opt}$, the contribution of the radiation backaction noise from the phase quadrature of the input vacuum noise in $R(\omega)$ can be totally eliminated.
Note that $A(\omega)>0$ and $\mathcal{K}(\omega)\geq0$, thus $\tan\phi^{opt}<0$, so the optimal phase $\phi^{opt}$ should be within the range $-\frac{\pi}{2}<\phi^{opt}\leq0$.

For convenience, $\mathcal{J}$ can be written as
\begin{eqnarray}\label{21}
\mathcal{J}=\frac{\hbar\eta^{2}c_{s}^{2}}{m\kappa_{0}}=\mathcal{J}_{0}\frac{1}{(1-\frac{2G}{\kappa_{0}})^{2}}.
\end{eqnarray}
 Without the PA ($G=0$), $\mathcal{J}=\mathcal{J}_{0}=\frac{\hbar\eta^{2}c_{s0}^{2}}{m\kappa_{0}}$, and $c_{s0}=\frac{\sqrt{2\kappa_{0}}\varepsilon_{l}}{\kappa_{0}}$ is the steady-state amplitude of the cavity field without the PA. In this case, $A(\omega)=1$, $\mathcal{K}(\omega)=\mathcal{J}_{0}\frac{\kappa_{0}^2}{\kappa_{0}^{2}+\omega^2}$, $\tan\phi^{opt}=-\mathcal{J}_{0}\frac{\kappa_{0}^2}{\kappa_{0}^{2}+\omega^2}$, $\mu(\omega)=\frac{1}{4\mathcal{J}_{0}}(1+\frac{\omega^{2}}{\kappa_{0}^{2}})$. Here $\mu(\omega)$ increases with $\omega$, but the detection frequency $\omega$ could not be zero, the minimum value of $\mu(\omega)$ is about $\frac{1}{4\mathcal{J}_{0}}$ when $\omega/\kappa_{0}\ll1$. The effect of the parameter $\mathcal{J}_{0}$ on the maximal force sensitivity for $G=0$ has been shown in \cite{Matsko}. Using the parameters $\eta=4.182\times10^{8}$ $\frac{1}{\text{m}}$,  $m=100$ ng, $\kappa_{0}=2\pi\times1$ MHz, the wavelength of the input laser $\lambda=1064$ nm, $\mathcal{J}_{0}=1/2$ requires rather large values of the driving powers $\wp=10$ W \cite{fn1}. Without the PA, the minimum value of $\mu(\omega)$ is about $\frac{1}{2}$, the force sensitivity is at the SQL. With improved dissipative systems, one should be able to reduce power levels as $\mathcal{J}_{0}$ scales as $\eta^{2}$.

 \begin{figure}[htp]
\begin{center}
\scalebox{0.6}{\includegraphics{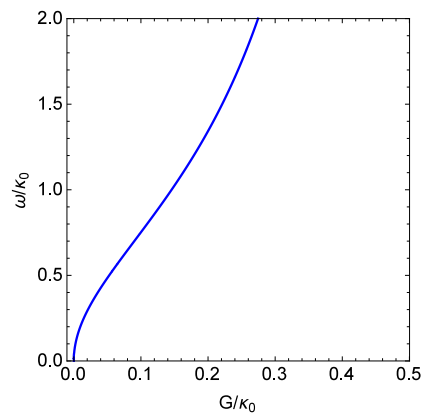}}
\caption{\label{Fig1} (Color online) The contour plot of $\mathcal{K}(\omega)=1/2$ as functions of the detection frequency $\omega/\kappa_{0}$ and the parametric gain $G/\kappa_{0}$ when $\mathcal{J}_{0}=1/2$.}
\end{center}
\end{figure}

The figure \ref{Fig1} shows the contour plot of $\mathcal{K}(\omega)=1/2$ as functions of the detection frequency $\omega/\kappa_{0}$ and the parametric gain $G/\kappa_{0}$ when $\mathcal{J}_{0}=1/2$. This curve represents the SQL since $\mu(\omega)=\frac{1}{4\mathcal{K}(\omega)}=1/2$. It is seen that the detection frequency $\omega$ increases with increasing the parametric gain $G$ of the PA. This is an advantage as one does not have to detect within the width of the resonator.

\begin{figure}[htp]
\begin{center}
\scalebox{0.85}{\includegraphics{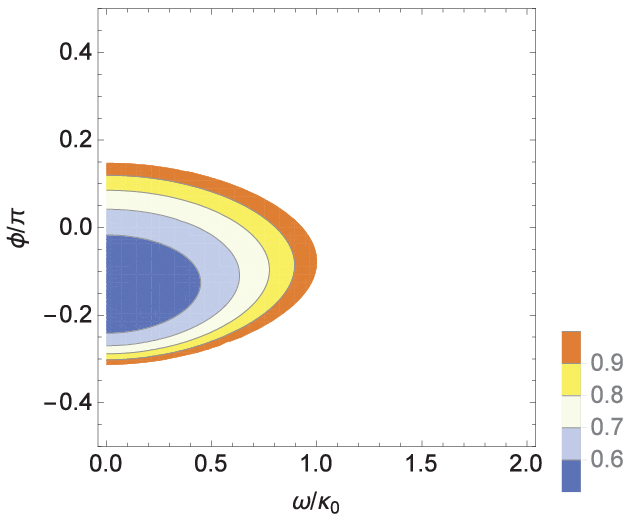}}
\scalebox{0.85}{\includegraphics{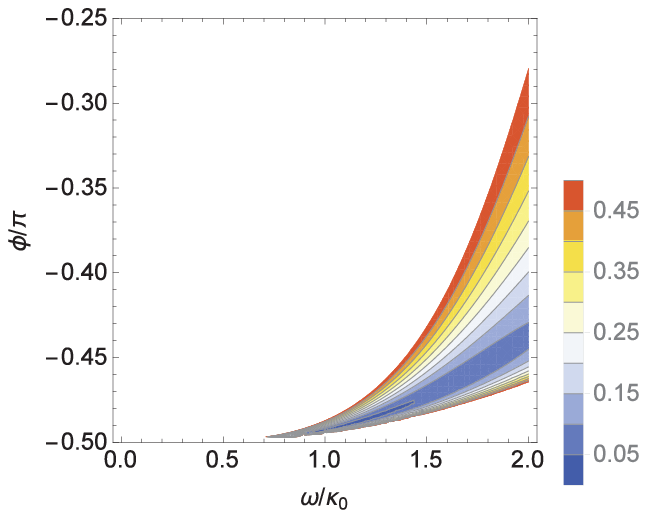}}
\caption{\label{Fig2} (Color online) The force sensitivity beyond the SQL: the contour plot of $\frac{R(\omega)}{F_{SQL}^{2}(\omega)}$ of the force measurement as functions of the detection phase $\phi/\pi$ and the detection frequency $\omega/\kappa_{0}$ for different parametric gains $G=0$ (upper), $G=0.4\kappa_{0}$ (lower) when $\mathcal{J}_{0}=1/2$.}
\end{center}
\end{figure}

The contour plot of $\frac{R(\omega)}{F_{SQL}^{2}(\omega)}$ as functions of the detection phase $\phi/\pi$ and the detection frequency $\omega/\kappa_{0}$ for different parametric gains  is shown in Fig. \ref{Fig2} when $\mathcal{J}_{0}=1/2$. For $G=0$, the minimum value of $\frac{R(\omega)}{F_{SQL}^{2}(\omega)}$ is in the range of 0.5--0.6. For $G=0.4\kappa_{0}$, the minimum value of $\frac{R(\omega)}{F_{SQL}^{2}(\omega)}$ can be less than 0.05. Thus the PA greatly improves the force sensitivity.

\begin{figure}[htp]
\begin{center}
\scalebox{0.85}{\includegraphics{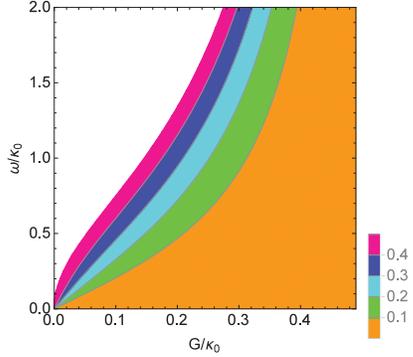}}
\caption{\label{Fig3} (Color online) The contour plot of $\mu(\omega)$ as functions of the detection frequency $\omega/\kappa_{0}$ and the parametric gain $G/\kappa_{0}$ when $\mathcal{J}_{0}=1/2$.}
\end{center}
\end{figure}

The figure \ref{Fig3} shows the contour plot of $\mu(\omega)$ as functions of the detection frequency $\omega/\kappa_{0}$ and the parametric gain $G/\kappa_{0}$ when $\mathcal{J}_{0}=1/2$. In the absence of the PA ($G=0$), the maximal force sensitivity is at the SQL when $\omega$ is close to zero. In the presence of the PA, increasing the parametric gain $G$ makes the maximal force sensitivity better than the SQL achievable in a larger frequency band. For example, when $0.28\leq G/\kappa_{0}<0.5$, to make the maximal force sensitivity $\mu(\omega)$ less than 0.5, the detection frequency can be in the range of $0<\omega/\kappa_{0}< 2$. Moreover, for $G\neq0$, we simplify Eq. (\ref{20}) as
\begin{eqnarray}\label{22}
\mu(\omega)=\frac{1}{4\mathcal{J}_{0}}\big(1-\frac{2G}{\kappa_{0}}\big)^{2}\big[(1+\frac{2G}{\kappa_{0}})^{2}+\frac{\omega^2}{\kappa_{0}^{2}}\big]\frac{\omega^{2}}{\omega^{2}+16G^{2}}.\nonumber\\
\end{eqnarray}
Since $G<\frac{\kappa_{0}}{2}$, when $\omega/G\ll1$, $\mu(\omega)\ll\frac{1}{4\mathcal{J}_{0}}$.
Thus the minimum value of $\mu(\omega)$ could be much smaller than that when there is no PA in the cavity. However, now the result is very sensitive to the damping and temperature of the particle. This would be discussed in Sec. III.

\section{Force Sensitivity of a free particle with damping and thermal fluctuations}
In practical situations, due to the interaction of the free particle with a thermal environment, we need to take into account the momentum damping  and the thermal noise of the free particle. In this case, the evolutions of the system operators become
\begin{eqnarray}\label{23}
\dot{q}&=&\frac{p}{m},\nonumber\\
\dot{p}&=&-\sqrt{2\kappa_{0}}\frac{\eta}{2}i\hbar[\varepsilon_{l}(c^{\dag}-c)+c^{\dag}c_{in}-c_{in}^{\dag}c]-\gamma_{m}p+\xi\nonumber\\
& &+f_{ex},\nonumber\\
\dot{c}&=&\sqrt{2\kappa_{0}}(1+\frac{\eta}{2}q)(\varepsilon_{l}+c_{in})+2Gc^{\dag}-\kappa_{0}(1+\eta q)c,\nonumber\\
\end{eqnarray}
where $\gamma_{m}$ is the mechanical damping rate of the free particle, $\xi$ is the thermal noise describing the coupling of the free particle to the thermal environment, it has zero mean value. It is assumed that the steady-state displacement $q_{s}$ of the free particle is zero. The steady-state mean values of the operators $p$ and $c$ are given by
\begin{eqnarray}\label{24}
p_{s}=0, \quad c_{s}=\frac{\sqrt{2\kappa_{0}}\varepsilon_{l}}{\kappa_{0}-2G}.
\end{eqnarray}
In the frequency domain, we obtain the fluctuation of the cavity field
\begin{eqnarray}\label{25}
& &\delta c(\omega)\nonumber\\
&=&\frac{1}{(\kappa_{0}-i\omega)^{2}-4G^{2}}\Big[-(\kappa_{0}-i\omega+2G)(\kappa_{0}+2G)\frac{\eta}{2}c_{s}\nonumber\\
& &\times\delta q(\omega)+\sqrt{2\kappa_{0}}(\kappa_{0}-i\omega)c_{in}(\omega)+2G\sqrt{2\kappa_{0}}c_{in}^{\dag}(-\omega)\Big],\nonumber\\
\end{eqnarray}
where the position fluctuation of the free particle is given by
\begin{eqnarray}\label{26}
\delta q(\omega)&=&\frac{i\hbar\eta\sqrt{\kappa_{0}}c_{s}}{\sqrt{2}m\omega(\omega+i\gamma_{m})}\frac{4G-i\omega}{\kappa_{0}-i\omega+2G}[c_{in}(\omega)\nonumber\\
& &-c_{in}^{\dag}(-\omega)]+\frac{\xi(\omega)+f_{ex}(\omega)}{-m\omega(\omega+i\gamma_{m})}.
\end{eqnarray}
Then using the input-output relation \cite{Walls}, we obtain the fluctuation of the output field
\begin{eqnarray}\label{27}
\delta c_{out}(\omega)&=&\sqrt{2\kappa_{0}}\frac{\eta}{2}c_{s}\frac{-i\omega-4G}{\kappa_{0}-i\omega-2G}\delta q(\omega)\nonumber\\
& &+\Big[\frac{2\kappa_{0}(\kappa_{0}-i\omega)}{(\kappa_{0}-i\omega)^{2}-4G^{2}}-1\Big]c_{in}(\omega)\nonumber\\
& &+\frac{4G\kappa_{0}}{(\kappa_{0}-i\omega)^{2}-4G^{2}}c_{in}^{\dag}(-\omega).
\end{eqnarray}
The amplitude of the output cavity field is found to be
\begin{eqnarray}\label{28}
\delta x_{out}(\omega)&=&\frac{\kappa_{0}+2G+i\omega}{\kappa_{0}-2G-i\omega}\Big[x_{in}(\omega)+\mathcal{K}_{n}(\omega)y_{in}(\omega)\nonumber\\
& &+B(\omega)u(\omega)\frac{\xi(\omega)}{F_{SQL}(\omega)}+B(\omega)u(\omega)\frac{f_{ex}(\omega)}{F_{SQL}(\omega)}\Big],\nonumber\\
\end{eqnarray}
where
\begin{eqnarray}\label{29}
B(\omega)&=&\frac{\sqrt{2\mathcal{K}_{n}(\omega)}}{\sqrt{1+i\frac{\gamma_{m}}{\omega}}},\nonumber\\
\mathcal{K}_{n}(\omega)&=&\mathcal{J}\frac{\kappa_{0}^{2}}{(\kappa_{0}+2G)^{2}+\omega^{2}}\frac{\omega^{2}+16G^{2}}{\omega(\omega+i\gamma_{m})}.
\end{eqnarray}
Here $\mathcal{K}_{n}(\omega)$ is dimensionless, and $\mathcal{J}$, $u(\omega)$, and $F_{SQL}(\omega)$ are the same as those in Eq. (\ref{11}).
Moreover, we obtain the phase quadrature of the output field
\begin{eqnarray}\label{30}
\delta y_{out}(\omega)&=&\frac{\kappa_{0}-2G+i\omega}{\kappa_{0}+2G-i\omega}y_{in}(\omega).
\end{eqnarray}
The generalized quadrature $\delta z_{out}(\omega)$ of the output field is given by
\begin{eqnarray}\label{31}
\delta z_{out}(\omega)&=&\frac{\kappa_{0}+2G+i\omega}{\kappa_{0}-2G-i\omega}\Big\{x_{in}(\omega)\cos\phi\nonumber\\
& &+\Big[\mathcal{K}_{n}(\omega)\cos\phi+\frac{1}{A(\omega)}\sin\phi\Big]y_{in}(\omega)\nonumber\\
& &+B(\omega)u(\omega)\frac{\xi(\omega)}{F_{SQL}(\omega)}\cos\phi\nonumber\\
& &+B(\omega)u(\omega)\frac{f_{ex}(\omega)}{F_{SQL}(\omega)}\cos\phi\Big\},
\end{eqnarray}
where $A(\omega)$ is the same as that in Eq. (\ref{13}). In Eq. (\ref{31}),
the first two terms are from the amplitude and phase quadratures of the input vacuum noise, respectively, the third term is from the thermal force $\xi(\omega)$, the last term is from the external force $f_{ex}(\omega)$.
Using the correlation functions of the input vacuum noise in Eq. (\ref{15}) and the correlation function of the thermal noise
\begin{eqnarray}\label{32}
\langle\xi(\omega)\xi(\Omega)\rangle=4\pi m k_{B}T\gamma_{m}\delta(\omega+\Omega),
\end{eqnarray}
where $k_{B}$ is the Boltzmann constant, $T$ is the temperature of the environment,
we obtain the spectrum of the output light
\begin{eqnarray}\label{33}
S_{zout}(\omega)&=&A(\omega)\Big\{\frac{1}{2}\cos^{2}\phi\nonumber\\
& &+\frac{1}{2}\Big|\mathcal{K}_{n}(\omega)\cos\phi+\frac{1}{A(\omega)}\sin\phi\Big|^{2}\nonumber\\
& &+|B(\omega)|^{2}\cos^{2}\phi\frac{2mk_{B}T\gamma_{m}}{F_{SQL}^{2}(\omega)}\nonumber\\
& &+|B(\omega)|^{2}\cos^{2}\phi\frac{S_{ex}(\omega)}{F_{SQL}^{2}(\omega)}\Big\}.
\end{eqnarray}
The force sensitivity of the system is determined by the quantity
\begin{eqnarray}\label{34}
R(\omega)&=&F_{SQL}^{2}(\omega)\Big\{\frac{1}{2|B(\omega)|^{2}}\nonumber\\
& &+\frac{1}{2|B(\omega)|^{2}}\big|\mathcal{K}_{n}(\omega)+\frac{1}{A(\omega)}\tan\phi\big|^{2}\nonumber\\
& &+\frac{k_{B}T}{\hbar\omega^{2}}\gamma_{m}\Big\}.
\end{eqnarray}
In Eq. (\ref{34}), the first term is the contribution of the photon shot noise from the amplitude quadrature of the input vacuum noise, the second term is the contribution of the radiation backaction noise from the phase quadrature of the input vacuum noise, the third term is the contribution of the thermal noise.
By choosing the optimal homodyne phase
\begin{eqnarray}\label{35}
\tan\phi^{opt}=A(\omega)\Big[-\frac{\mathcal{K}_{n}(\omega)\omega}{\omega-i\gamma_{m}}\Big],
\end{eqnarray}
where $\tan\phi^{opt}$ is real although the right-hand side of Eq. (\ref{35}) looks complex,
$R(\omega)$ takes the minimum value
\begin{eqnarray}\label{36}
R(\omega)_{min}&=&F_{SQL}^{2}(\omega)\Big\{\frac{1}{2|B(\omega)|^{2}}\nonumber\\
& &+\frac{1}{2|B(\omega)|^{2}}\mathcal{K}_{n}(\omega)\mathcal{K}_{n}(-\omega)\frac{\gamma_{m}^{2}}{\omega^{2}+\gamma_{m}^{2}}\nonumber\\
& &+\frac{k_{B}T}{\hbar\omega^{2}}\gamma_{m}\Big\}.
\end{eqnarray}
In Eq. (\ref{36}), the second term is the contribution of the radiation backaction noise, which is proportional to $\gamma_{m}^{2}$. Hence, when the mechanical damping is considered, the contribution of the radiation backaction noise can not be totally removed by choosing the optimal homodyne phase.
We take $\mathcal{J}_{0}=1/2$, and $\gamma_{m}/\kappa_{0}=10^{-5}$.

\subsection{Force sensitivity at zero temperature but with damping included}
First, we show how the mechanical damping affects the force sensitivity when $T=0$ K.
\begin{figure}[htp]
\begin{center}
\scalebox{0.85}{\includegraphics{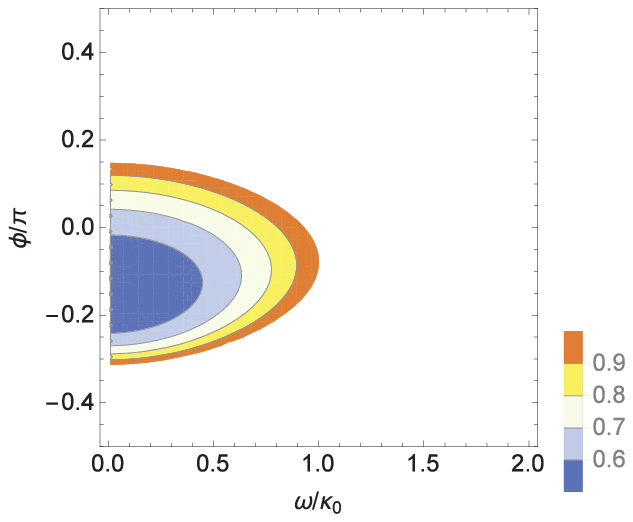}}
\scalebox{0.85}{\includegraphics{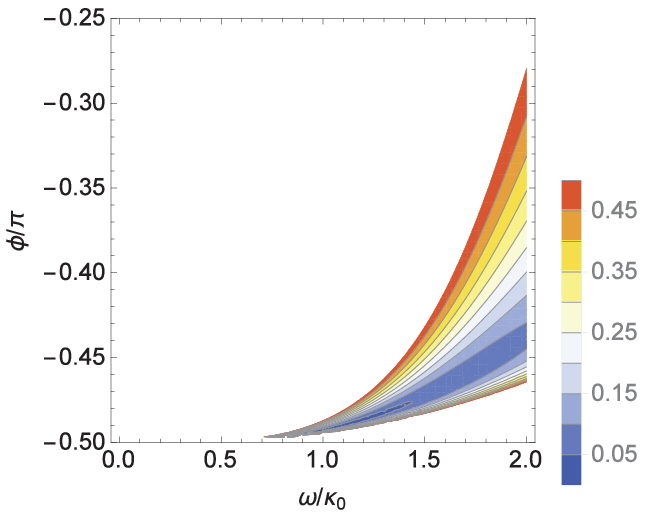}}
\caption{\label{Fig4} (Color online) The contour plot of $\frac{R(\omega)}{F_{SQL}^{2}(\omega)}$ of the force measurement as functions of the detection phase $\phi/\pi$ and the detection frequency $\omega/\kappa_{0}$ for different parametric gains $G=0$ (upper), $G=0.4\kappa_{0}$ (lower) when $\mathcal{J}_{0}=1/2$, $\gamma_{m}/\kappa_{0}=10^{-5}$, and $T=0$ K. Interestingly enough increasing $\gamma_{m}/\kappa_{0}$ to a value $10^{-2}$ does not produce any noticeable change in this figure.}
\end{center}
\end{figure}

 The contour plot of $\frac{R(\omega)}{F_{SQL}^{2}(\omega)}$ as functions of the detection phase $\phi/\pi$ and the detection frequency $\omega/\kappa_{0}$ for different parametric gains is shown in Fig. \ref{Fig4} when $\mathcal{J}_{0}=1/2$, $\gamma_{m}/\kappa_{0}=10^{-5}$, and $T=0$ K. For $G=0$, both the minimum values of  $\frac{R(\omega)}{F_{SQL}^{2}(\omega)}$ in Fig. \ref{Fig4} and Fig. \ref{Fig2} are in the range of 0.5--0.6. For $G=0.4\kappa_{0}$, both the minimum values of  $\frac{R(\omega)}{F_{SQL}^{2}(\omega)}$ in Fig. \ref{Fig4} and Fig. \ref{Fig2} are less than 0.05. Hence, there is no apparent difference between the minimum values of $\frac{R(\omega)}{F_{SQL}^{2}(\omega)}$ with and without the mechanical damping, thus the mechanical damping has no apparent effect on the force sensitivity in the absence or presence of the PA.

\subsection{Robustness of the force sensitivity beyond the SQL against the effect of the temperature}
Next, we show how the thermal noise of the free particle affects the force sensitivity.

\begin{figure}[htp]
\begin{center}
\scalebox{0.85}{\includegraphics{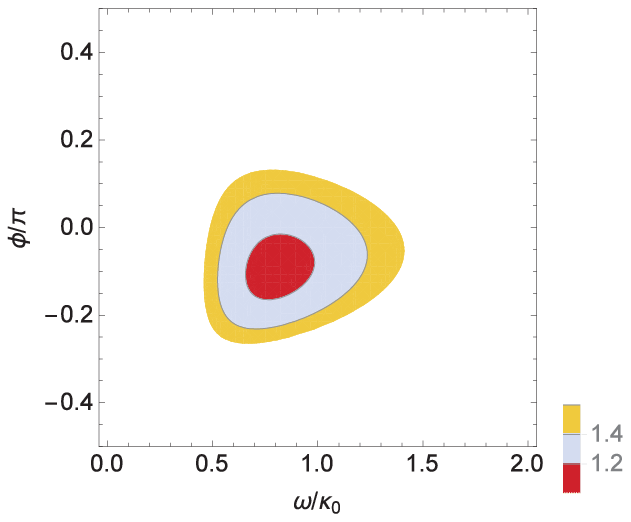}}
\scalebox{0.85}{\includegraphics{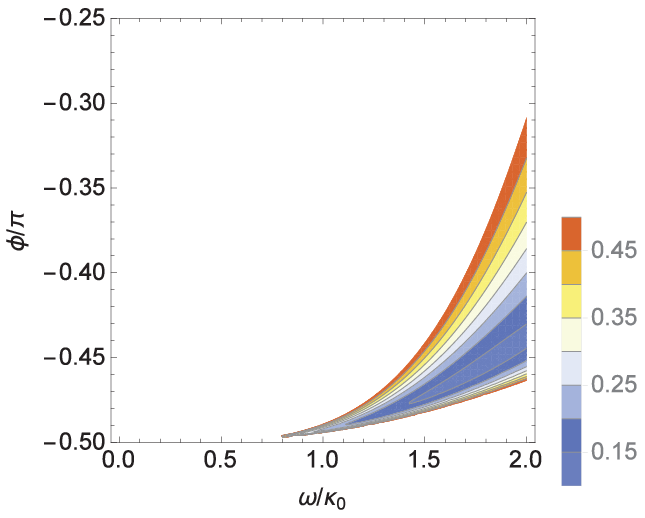}}
\caption{\label{Fig5} (Color online) The contour plot of $\frac{R(\omega)}{F_{SQL}^{2}(\omega)}$ of the force measurement as functions of the detection phase $\phi/\pi$ and the detection frequency $\omega/\kappa_{0}$ for different parametric gains $G=0$ (upper), $G=0.4\kappa_{0}$ (lower) when $\mathcal{J}_{0}=1/2$, $\gamma_{m}/\kappa_{0}=10^{-5}$, $\kappa_{0}=2\pi\times1$ MHz, and $T=1$ K.}
\end{center}
\end{figure}

\begin{figure}[htp]
\begin{center}
\scalebox{0.85}{\includegraphics{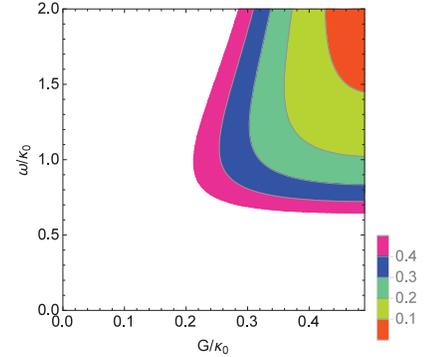}}
\caption{\label{Fig6} (Color online) The contour plot of $\mu(\omega)$ as functions of the detection frequency $\omega/\kappa_{0}$ and the parametric gain $G/\kappa_{0}$ when $\mathcal{J}_{0}=1/2$, $\gamma_{m}/\kappa_{0}=10^{-5}$, $\kappa_{0}=2\pi\times1$ MHz, and $T=1$ K.}
\end{center}
\end{figure}

The contour plot of $\frac{R(\omega)}{F_{SQL}^{2}(\omega)}$ as functions of the detection phase $\phi/\pi$ and the detection frequency $\omega/\kappa_{0}$ for different parametric gains is shown in Fig. \ref{Fig5} when $\mathcal{J}_{0}=1/2$, $\gamma_{m}/\kappa_{0}=10^{-5}$, $\kappa_{0}=2\pi\times1$ MHz, and $T=1$ K. For $G=0$, the minimum value of $\frac{R(\omega)}{F_{SQL}^{2}(\omega)}$ in Fig. \ref{Fig5} ($T=1$ K) is in the range of 1.0--1.2, while the minimum value of $\frac{R(\omega)}{F_{SQL}^{2}(\omega)}$ in Fig. \ref{Fig4} ($T=0$ K) is in the range of 0.5--0.6. Hence the thermal noise decreases the force sensitivity significantly in the absence of the PA, and is above the SQL. For $G=0.4\kappa_{0}$, the minimum value of $\frac{R(\omega)}{F_{SQL}^{2}(\omega)}$ in Fig. \ref{Fig5} ($T=1$ K) is in the range of 0.1--0.15, while the minimum value of $\frac{R(\omega)}{F_{SQL}^{2}(\omega)}$ in Fig. \ref{Fig4} ($T=0$ K) is less than 0.05. Thus the thermal noise reduces the force sensitivity in the presence of the PA with $G=0.4\kappa_0$. Therefore, the influence of the thermal noise on the force sensitivity of the system becomes smaller in the presence of the PA with a higher parametric gain.

The figure \ref{Fig6} shows the contour plot of $\mu(\omega)$ as functions of the detection frequency $\omega/\kappa_{0}$ and the parametric gain $G/\kappa_{0}$ when $\mathcal{J}_{0}=1/2$, $\gamma_{m}/\kappa_{0}=10^{-5}$, $\kappa_{0}=2\pi\times1$ MHz, and $T=1$ K. In Fig. \ref{Fig6} ($T=1$ K), it is seen that the maximal force sensitivity $\mu(\omega)$ less than 0.1 is possible to achieve when $0.426\leq G/\kappa_{0}<0.5$ and $1.45\leq\omega/\kappa_{0}<2$. In Fig. \ref{Fig3} ($\gamma_{m}=0$ and $T=0$ K), we note that the maximal force sensitivity $\mu(\omega)$ less than 0.1 exists when $0< G/\kappa_{0}<0.5$ and $0<\omega/\kappa_{0}<2$. Therefore, the presence of the PA with a higher parametric gain makes the maximal force sensitivity of the system become more robust against the thermal environment. Furthermore,
from Eq. (\ref{36}), the maximal force sensitivity $\mu(\omega)$ can be expressed as
\begin{eqnarray}\label{37}
\mu(\omega)&=&\frac{1}{4\mathcal{J}_{0}}\big(1-\frac{2G}{\kappa_{0}}\big)^{2}\frac{(\kappa_{0}+2G)^{2}+\omega^{2}}{\kappa_{0}^{2}}\frac{\omega^{2}+\gamma_{m}^{2}}{\omega^{2}+16G^{2}}\nonumber\\
& &+\frac{\mathcal{J}_{0}}{4}\frac{1}{\big(1-\frac{2G}{\kappa_{0}}\big)^{2}}\frac{\kappa_{0}^{2}}{(\kappa_{0}+2G)^{2}+\omega^{2}}\frac{\omega^{2}+16G^{2}}{\omega^{2}+\gamma_{m}^{2}}\frac{\gamma_{m}^{2}}{\omega^{2}}\nonumber\\
& &+\frac{k_{B}T}{\hbar\omega^{2}}\gamma_{m}.
\end{eqnarray}
For $G=0$, Eq. (\ref{37}) becomes
\begin{eqnarray}\label{38}
\mu(\omega)&=&\frac{1}{4\mathcal{J}_{0}}\frac{\kappa_{0}^{2}+\omega^{2}}{\kappa_{0}^{2}}\frac{\omega^{2}+\gamma_{m}^{2}}{\omega^{2}}+\frac{\mathcal{J}_{0}}{4}\frac{\kappa_{0}^{2}}{\kappa_{0}^{2}+\omega^{2}}\frac{\omega^{2}}{\omega^{2}+\gamma_{m}^{2}}\frac{\gamma_{m}^{2}}{\omega^{2}}\nonumber\\
& &+\frac{k_{B}T}{\hbar\omega^{2}}\gamma_{m}.
\end{eqnarray}
In view of the complexity of these expressions, we present in the table I, the minimum value of the parameter $\mu(\omega)$ for  $\gamma_{m}/\kappa_{0}=10^{-5}$, $\kappa_{0}=2\pi\times1$ MHz, and $\mathcal{J}_{0}=1/2$, $1/10$, $1/50$. The great advantage of using the PA is obvious. For $\mathcal{J}_{0}=1/2$ and $T=1$ K, the PA can enhance the force sensitivity by a factor of about 7 over the SQL. For $T=1$ K, if we decrease the input laser power so that the parameter $\mathcal{J}_{0}$ is $\frac{1}{10}$ (power $\wp=2$ W) or $\frac{1}{50}$ (power $\wp=0.4$ W), it is seen that the PA can still make the force sensitivity below the SQL. Remarkably, the table I shows that for a power level 0.4 W, we still can get an improvement by a factor of about 2 below the SQL.
As noted in \cite{Peano} the mechanical damping is a critical parameter, we have checked how the results in the table I are affected by increasing $\gamma_{m}$ by two orders of magnitude. Interestingly enough as seen from the table II that at 10 mK one can still get an improvement by a factor of 2 over the SQL for the power $\wp=0.4$ W.

\begin{table}[h]
\begin{tabular}{|c|c|c|c|c|c|}
 \hline
 \multirow{4}{*}{$\mathcal{J}_{0}=1/2$}& $T$ &\multicolumn{2}{|c|}{0 K}&  \multicolumn{2}{|c|}{1 K}\\
 \cline{2-6}
&$G/\kappa_{0}$ &0 &0.46&0 &0.46 \\
\cline{2-6}
&$\omega/\kappa_{0}$&0.003 & 0.1 &0.8& 1.9 \\
\cline{2-6}
&$\mu(\omega)_{min}$& 0.5 & $5.28\times10^{-5}$ & 1.14 & 0.07\\
 \hline
\multirow{4}{*}{$\mathcal{J}_{0}=1/10$}& $T$ &\multicolumn{2}{|c|}{0 K}&  \multicolumn{2}{|c|}{1 K}\\
 \cline{2-6}
&$G/\kappa_{0}$ &0 &0.46&0 &0.46 \\
\cline{2-6}
&$\omega/\kappa_{0}$&0.003 & 0.06 &0.5& 1.9 \\
\cline{2-6}
&$\mu(\omega)_{min}$& 2.5 & $1\times10^{-4}$ & 3.96 & 0.118\\
 \hline
 \multirow{4}{*}{$\mathcal{J}_{0}=1/50$}& $T$ &\multicolumn{2}{|c|}{0 K}&  \multicolumn{2}{|c|}{1 K}\\
 \cline{2-6}
&$G/\kappa_{0}$ &0 &0.46&0 &0.46 \\
\cline{2-6}
&$\omega/\kappa_{0}$&0.003 & 0.034 &0.4& 1.3 \\
\cline{2-6}
&$\mu(\omega)_{min}$& 12.5 & $1.54\times10^{-4}$ & 15.8 & 0.266\\
 \hline
\end{tabular}
\caption{The minimum value of $\mu(\omega)$ for $\gamma_{m}/\kappa_{0}=10^{-5}$, $\kappa_{0}=2\pi\times1$ MHz, $\mathcal{J}_{0}=1/2$, $1/10$, $1/50$, and the corresponding input laser power $\wp=10$ W, 2 W, 0.4 W.}
\end{table}

\begin{table}[h]
\begin{tabular}{|c|c|c|c|}
 \hline
\multirow{4}{*}{$\mathcal{J}_{0}=1/10$}& $T$ &  \multicolumn{2}{|c|}{10 mK}\\
 \cline{2-4}
&$G/\kappa_{0}$ &0 &0.46 \\
\cline{2-4}
&$\omega/\kappa_{0}$&0.5& 1.9 \\
\cline{2-4}
&$\mu(\omega)_{min}$& 3.96 & 0.118\\
 \hline
 \multirow{4}{*}{$\mathcal{J}_{0}=1/50$}& $T$ & \multicolumn{2}{|c|}{10 mK}\\
 \cline{2-4}
&$G/\kappa_{0}$&0 &0.46 \\
\cline{2-4}
&$\omega/\kappa_{0}$&0.36& 1.3 \\
\cline{2-4}
&$\mu(\omega)_{min}$& 15.73 & 0.266\\
 \hline
\end{tabular}
\caption{The minimum value of $\mu(\omega)$ for $\gamma_{m}/\kappa_{0}=10^{-3}$, $\kappa_{0}=2\pi\times1$ MHz, $\mathcal{J}_{0}=1/10$, $1/50$, and the corresponding input laser power $\wp=2$ W, 0.4 W.}
\end{table}

\section{Conclusions}
In conclusion, we have investigated the sensitivity of the force detection of the dissipative system with a PA. It is found that the PA in the dissipative system makes the maximal force sensitivity better than the SQL achievable in a wide frequency band. Meanwhile, it is found that the PA largely enhances the maximal force sensitivity. Moreover, increasing the parametric gain $G$ of the PA reduces the influences of the mechanical damping and the thermal noise, resulting in a better force sensitivity. The PA can improve the maximal force sensitivity by a factor of about 7 for the temperature of the environment $T=1$ K. This is for the power level corresponding to the SQL when $G=0$. The result obtained above is useful for ultrasensitive detection of weak forces based on nanomechanical systems.

\setcounter{equation}{0}  
\section*{APPENDIX: THE COMPARISON OF THE FORCE SENSITIVITY OF A MECHANICAL OSCILLATOR WITH A FREE PARTICLE}
\renewcommand{\theequation}{A\arabic{equation}}
If the free particle is replaced by a mechanical oscillator with effective mass $m$ and resonance frequency $\omega_{m}$,
 the Hamiltonian of the system in the rotating frame at the input laser frequency $\omega_{l}$ becomes
\begin{eqnarray}\label{A1}
H&=&\hbar(\omega_c-\omega_l)c^{\dag}c+\frac{p^{2}}{2m}+\frac{1}{2}m\omega_{m}^{2}q^{2}\nonumber\\
& &+i\hbar\sqrt{2\kappa(q)}[\varepsilon_{l}(c^{\dag}-c)+c^{\dag}c_{in}-c_{in}^{\dag}c]\nonumber\\
& &+i\hbar G(c^{\dag2}-c^{2}),
\end{eqnarray}
where $q$ and $p$ are the position and momentum operators of the mechanical oscillator, respectively, they satisfy $[q,p]=i\hbar$.
Through calculations, we find the force sensitivity of the mechanical oscillator
\begin{eqnarray}\label{A2}
R_{mo}(\omega)&=&F_{moSQL}^{2}\frac{1}{u_{mo}(\omega)u_{mo}(-\omega)}\frac{1}{4\mathcal{K}_{mo}(\omega)}\nonumber\\
& &\times\left\{1+\Big[\mathcal{K}_{mo}(\omega)+\frac{1}{A(\omega)}\tan\phi\Big]^{2}\right\},
\end{eqnarray}
where
\begin{eqnarray}
& &\mathcal{K}_{mo}(\omega)=\mathcal{J}\frac{\kappa_{0}^{2}}{(\kappa_{0}+2G)^{2}+\omega^{2}}\frac{\omega^{2}+16G^{2}}{\omega^{2}-\omega_{m}^{2}},\nonumber\\
& &\mathcal{J}=\frac{\hbar\eta^{2}c_{s}^{2}}{m\kappa_{0}},\nonumber\\
& &F_{moSQL}=\sqrt{2m\hbar\omega_{m}^{2}},\nonumber\\
& &u_{mo}(\omega)=\frac{\sqrt{16G^{2}+\omega^{2}}}{\sqrt{(\kappa_{0}+2G)^{2}+\omega^{2}}}\frac{\kappa_{0}+2G-i\omega}{\omega+4Gi}i\nonumber\\
& &\qquad\qquad\quad\times\frac{\omega_{m}}{\sqrt{\omega^{2}-\omega_{m}^{2}}},\nonumber\\
& &u_{mo}(\omega)u_{mo}(-\omega)=\frac{\omega_{m}^{2}}{\omega^{2}-\omega_{m}^{2}},\nonumber\\
& &A(\omega)=\frac{(\kappa_{0}+2G)^{2}+\omega^{2}}{(\kappa_{0}-2G)^{2}+\omega^{2}}.
\end{eqnarray}
When the second term in $R_{mo}(\omega)$ is zero, we obtain the optimal phase $\tan\phi^{opt}=-A(\omega)\mathcal{K}_{mo}(\omega)$, the minimum value of $R_{mo}(\omega)$ is found to be
\begin{eqnarray}\label{A3}
R_{mo}(\omega)_{min}&=&F_{moSQL}^{2}\frac{1}{u_{mo}(\omega)u_{mo}(-\omega)}\frac{1}{4\mathcal{K}_{mo}(\omega)}.\nonumber\\
\end{eqnarray}
We have shown that the force sensitivity of the free particle is given by
\begin{eqnarray}\label{A4}
R(\omega)&=&F_{SQL}^{2}(\omega)\frac{1}{4\mathcal{K}(\omega)}\left\{1+\Big[\mathcal{K}(\omega)+\frac{1}{A(\omega)}\tan\phi\Big]^{2}\right\},\nonumber\\
\end{eqnarray}
where
\begin{eqnarray}
& &\mathcal{K}(\omega)=\mathcal{J}\frac{\kappa_{0}^{2}}{(\kappa_{0}+2G)^{2}+\omega^{2}}\frac{\omega^{2}+16G^{2}}{\omega^{2}},\nonumber\\
& &\mathcal{J}=\frac{\hbar\eta^{2}c_{s}^{2}}{m\kappa_{0}},\nonumber\\
& &F_{SQL}(\omega)=\sqrt{2m\hbar\omega^{2}},\nonumber\\
& &A(\omega)=\frac{(\kappa_{0}+2G)^{2}+\omega^{2}}{(\kappa_{0}-2G)^{2}+\omega^{2}}.
\end{eqnarray}
When the second term in $R(\omega)$ is zero, we obtain the optimal phase
$\tan\phi^{opt}=-A(\omega)\mathcal{K}(\omega)$,
the minimum value of $R(\omega)$ is found to be
\begin{eqnarray}\label{A5}
R(\omega)_{min}=\frac{F_{SQL}^{2}(\omega)}{4\mathcal{K}(\omega)}.
\end{eqnarray}
It is noted that
\begin{eqnarray}
\frac{R_{mo}(\omega)_{min}}{R(\omega)_{min}}=(1-\frac{\omega_{m}^{2}}{\omega^{2}})^{2}.
\end{eqnarray}
If $\omega_{m}\ll\omega$, we obtain $\frac{R_{mo}(\omega)_{min}}{R(\omega)_{min}}\approx1$.
If we choose $\omega_{m}\ll\kappa_{0}$, $\omega\gg\omega_{m}$, but $\omega\lesssim 2\kappa_{0}$, then $\frac{R_{mo}(\omega)_{min}}{R(\omega)_{min}}\approx1$.


\begin{thebibliography}{99}
\bibitem{fn} An analysis of the position sensitivity using a PA in a dispersive optomechanical system has been presented in \cite{Peano}. These authors point out the great advantage of the PA in allowing one to use mechanical resonators with much larger decay rates.
\bibitem{fn1} In Ref. \cite{Hammerer3}, the dissipative coupling strength $g_{k}$ can be $2\pi\times 0.25$ Hz, it is defined as $g_{k}=\eta \kappa_{0}\sqrt{\frac{\hbar}{2m\omega_{m}}}$. When the mass of the mechanical membrane is $m=100$ ng, the cavity decay rate in the absence of the dissipative coupling is $\kappa_{0}=2\pi\times1$ MHz, the resonance frequency of the membrane is $\omega_{m}=2\pi\times 234.8$ kHz, the dissipative coupling constant is found to be $\eta=4.182\times10^{8}\ \frac{1}{\text{m}}$.
\end{thebibliography}
\end{document}